\documentclass[twocolumn,aps,showpacs,prl]{revtex4}
\usepackage{graphicx}
\usepackage{dcolumn}
\usepackage{bm}

\def \ba122{BaFe$_2$As$_2$}
\def \sr122{SrFe$_2$As$_2$}

\def \hc2{$\mu_0H_{c2}$}

\begin{document}

\title{Quantum oscillations in the parent pnictide BaFe$_2$As$_2$ : itinerant electrons in the reconstructed state}

\author{James G. Analytis$^1$, Ross D. McDonald$^2$, Jiun-Haw Chu$^1$, Scott C. Riggs$^2$,   Alimamy F. Bangura$^3$, Chris Kucharczyk$^1$, Michelle Johannes$^4$, I. R. Fisher$^1$}
  \affiliation{$^1$Geballe Laboratory for Advanced Materials and
  Department of Applied Physics, Stanford University, CA 94305, USA}
\affiliation{$^2$Los Alamos National Laboratory, Los Alamos, NM, USA}
\affiliation{$^3$H. H. Wills Physics Laboratory, University of Bristol, 1 Tyndall Ave., Bristol BS8 1TL, UK}
\affiliation{$^4$Center for Computational Materials Science, Naval Research Laboratory Laboratory, Washington, D.C. 20375, USA}

\date{\today}

\begin{abstract} 
We report quantum oscillation measurements that enable the direct observation of the Fermi surface of the low temperature ground state of \ba122. From these measurements we characterize the low energy excitations, revealing that the Fermi surface is reconstructed in the antiferromagnetic state, but leaving itinerant electrons in its wake. The present measurements are consistent with a conventional band folding picture of the antiferromagnetic ground state, placing important limits on the topology and size of the Fermi surface. 
\end{abstract}



\maketitle

The nature of superconductivity in Fe-pnictide family of compounds has thus far eluded a universally accepted explanation. Part of the problem is understanding the fundamental quasiparticle dynamics of the parent compounds which show evidence for electron itineracy on the one hand\cite{coldea,sebastian,hsieh,digiorgi}, and local magnetism on the other\cite{yang,delacruz,ewings,zhao}. Recent ARPES measurements have suggested a novel exchange mechanism driving the magnetism\cite{yang,zhang} due to an apparent band splitting at the transition temperature T$_{SDW}$ (135K in \ba122) while other ARPES measurements have suggested Fermi surface nesting\cite{hsieh,zabolotnyy} by mapping the shape of the observed Fermi pockets to an inferred nesting instability. In addition, while neutron data has suggested the complete suppression of magnetic order in F-doped CeFeAsO before the material becomes superconducting\cite{delacruz}, muon spectroscopy has detected magnetic fluctuations inside the superconducting dome\cite{blundell} in F-doped SmFeAsO. ARPES has also observed the persistence of nesting instabilities in the superconducting state of K-doped \ba122\cite{zabolotnyy}. These observations have left open such questions as to what the role of disorder and magnetism is in shaping the superconducting mechanism, whether the superconducitvity emerges from the normal state Fermi surface or the reconstructed state, or even what the microscopic nature of the magnetism is in the parent compounds\cite{mazinjohannes}. Resolving these issues requires that the low energy quasiparticle excitations are revealed. This is especially true because knowledge of the itinerant nature of the low temperature ground state places significant constraints on the magnetism associated with the order. In the present paper we report quantum oscillation (QO) measurements consistent with a nesting mechanism that folds bands of the non-magnetic state in a conventional manner. As predicted by recent theoretical investigations, we find that the SDW instability does not fully gap the Fermi surface\cite{ran}.

In the measurements reported here on \ba122\, we use two separate techniques, torque magnetometry and a radio frequency contactless conductivity technique using a tunnel diode oscillator (TDO), both of which have been used recently to observe oscillations in the closely related compounds LaFePO\cite{coldea} and \sr122.\cite{sebastian} We observe three small pockets comprising 1.7$\%$, 0.7$\%$ and 0.3$\%$, of the paramagnetic Brillouin zone
(that associated with the tetragonal state) and produce band structure calculations of a reconstructed state which are in broad agreement. Furthermore we map the topology of these small pockets and extract their effective mass. The present measurements illustrate that itinerant electrons play a fundamental role in the ordered state of ternary Fe-pnictides, and place important limits on the topology and size of the Fermi surface in the antiferromagnetic state. 
\begin{figure}[ht]
\includegraphics[width=7.2cm]{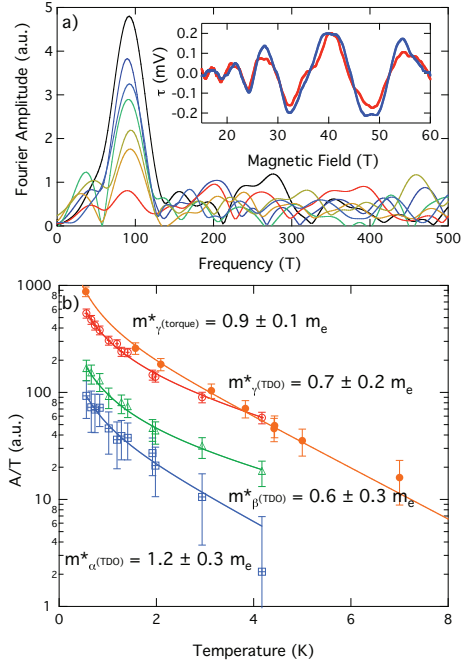}
\caption{(Color online) a) Thermal evolution of the Fourier spectrum of the torque data measured with the applied magnetic field oriented 27$^\circ$ from the c-axis. The corresponding temperatures from highest to lowest intensity are 0.5, 2, 3.1, 4.0, 6.0, 8.0 and 9.0~K. The inset shows typical low-temperature (0.5~K and 1.5~K) torque data (with background subtracted). b) The temperature dependence of the Fourier amplitude (A) for the $\alpha$ pocket (open squares), the $\beta$ pocket (open triangles), the $\gamma$ pocket (open circles) all measured by the TDO technique and the $\gamma$ pocket as measured by torque magnetometry (solid circles). The solid lines are fits to the LK formula yielding the following effective masses (at $\theta=23^\circ$); $m^{\star}_{\alpha} (TDO) = 1.2\pm0.3~m_e$, $m^{\star}_{\beta} (TDO) = 0.6\pm0.3~m_e$, $m^{\star}_{\gamma} (TDO) = 0.7\pm0.2~m_e$ and $m^{\star}_{\gamma} (Torque) = 0.9\pm0.1~m_e$.} \label{data}
\end{figure}

Single crystal samples of \ba122\, were prepared by slow cooling a
ternary melt, as described elsewhere.\cite{jiunhaw,mandrus}. The
crystals were then annealed at high temperature $\sim$ 900 $^\circ$C
in vacuum for 24hrs to allow interstitial and vacancy disorder to
relax. Sample surfaces often appeared degraded after the annealing
process, but residual resistivity ratios increased from 4 to 10. For these crystals the
absolute values of the in-plane resistivity are around $\sim
0.7$m$\Omega$\,cm at 300K. Magnetic QO measurements were performed at
the National High Magnetic Field Laboratory (Los Alamos) in the short
pulse ($\sim$10ms rise time) 65T and long pulse ($\sim$1s rise time) 60T magnets. For the torque magnetometry experiments, piezoresistive microcantilevers were used at
temperatures down to 0.4 K. The measured torque signal is dependent on the anisotropic magnetization of the sample $\tau\propto\mu_0{\mathbf M}\times{\mathbf H}$, and can thus detect magnetic field dependent oscillations in the magnetisation, known as the Haas-van Alphen effect. Sample A, used on the torque cantilever was 0.2$\times$0.2$\times$0.08
mm$^3$. Another sample (sample B) of dimensions 2$\times$2$\times$0.3mm$^3$ was mounted with its tetragonal $c$ axis parallel to the axis of a compensated coil that forms part of the tunnel diode oscillator circuit. The oscillator resonates at frequency $\sim$
37 MHz in the absence of an applied field, dropping by $\sim$ 300 kHz at 65T in response to the magnetoconductivity of the sample. As the skin depth changes due to the Shubnikov-de Haas effect, the coil resonance frequency is correspondingly altered.

Background subtracted data taken from sample A is shown in the inset of Figure \ref{data} (a) (a smooth polynomial of order 3). The Fourier content of the data is shown in Figure \ref{data} (a), illustrating the predominance of a single frequency F$_\gamma=80$ (herein the $\gamma$ pocket, which appears at 95T in Figure \ref{data} because the angle between the field and the tetragonal $c$-axis $\theta=27^\circ$). In sample B two higher frequencies F$_\alpha$=440, F$_\beta$=190 T appear at $\theta=0^\circ$ which we shall call the $\alpha$ and $\beta$ pockets respectively. The $\alpha$, $\beta$ and $\gamma$ orbits comprise about 1.7$\%$, 0.7$\%$ and 0.3$\%$ of the paramagnetic Brillouin zone.

\begin{figure}[ht]
\includegraphics[width=7.2cm]{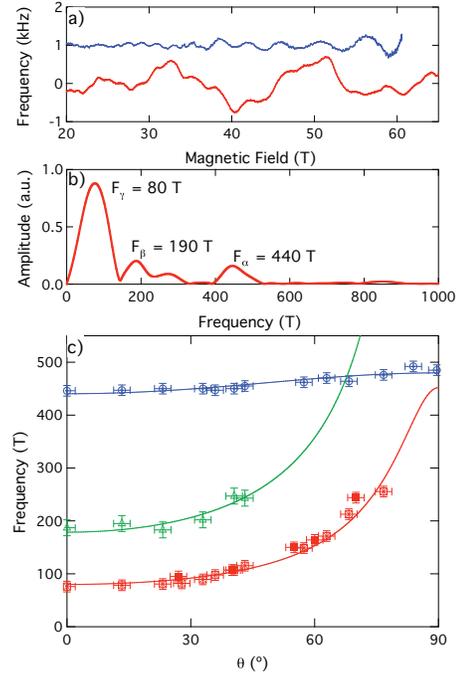}
\caption{(Color online) a) The residual frequency shift of the TDO circuit (once a 3rd order background has been subtracted) showing QOs periodic in inverse magnetic field. The upper curve is for the magnetic field oriented parallel to the c-axis $\theta=0^\circ$ and the lower curve for field perpendicular to the c-axis. b) The Fourier spectrum of the TDO data at $\theta=0^\circ$. c) The field orientation dependence of the QO frequencies. The hollow points are extracted form the TDO data the solid points are extracted from the torque data. The solid lines are fits to Fermi surface volumes with an elliptical cross section oriented parallel to the c-axis. The $\alpha$, $\beta$ and $\gamma$ frequencies correspond to a Fermi surface cross-section comprising 1.7$\pm0.05\%$, 0.7$\pm0.05\%$ and 0.30$\pm0.02\%$  of the paramagnetic Brillouin zone. The ellipticity for $\alpha$ is $1.1\pm.1$, $\beta$ is $5\pm1$ and $\gamma$ is $5.6\pm.1$.}
\label{freqmass}
\end{figure}

We extract the effective mass by fitting the temperature dependence of the oscillation amplitude with the thermal damping term R$_T=X/$sinh$(X)$ of the Lifshitz-Kosevich(LK) formalism, where X=14.69$m^*T/B$ and $m^*$ is the effective mass. Presently $1/B$ is the average inverse field of the Fourier window, taken between 20 and 60 T\cite{shoenberg}. The suppression of the $\alpha$, $\beta$ and $\gamma$ frequency amplitudes is shown in Figure \ref{data} (b). Data for the $\gamma$ pocket are shown for both techniques. The effective mass is $m_\gamma^*=0.9\pm 0.1m_e$ in sample A and $m_\gamma^*=0.7\pm0.2m_e$ in sample B which are in broad agreement. The $\alpha$ pocket has a mass of $m_\alpha^*=1.2\pm 0.3$$m_e$, the $\beta$ pocket has a mass of $m_\beta^*=0.6\pm 0.3$$m_e$. The errors given in Figure \ref{freqmass} (c) are determined by the noise floor of the Fourier spectra. Furthermore we estimate the Dingle temperature for the pockets to be $T_D^\alpha=$4K$\pm1$ and $T_D^\gamma=$3K$\pm1$ though we are unable to extract this for the $\beta$ pocket. The Dingle temperature was not accounted for in our effective mass fitting\cite{errorfootnote}.

We next turn to the angle dependence of the observed pockets. Sample B was mounted on a probe that could be continuously rotated allowing us to collect a comprehensive data set on all frequencies. Sample A was discretely rotated and the QOs were only observable between($27^\circ<\theta<70^\circ$), due to the loss in torque signal as the orientation of the field approaches a crystal symmetry direction. Even in sample B the intensity of $\gamma$ QOs is lost at angles $\theta>80^\circ$ and $\beta$ for $\theta>50^\circ$. The $\gamma$ and $\beta$ pockets are highly eccentric, resembling elongated cigar shapes. By contrast, the $\alpha$ pocket has a very small angular dependence, suggesting that the pocket is much more isotropic and three-dimensional. This frequency is likely not observable in the torque technique due to the smaller signal/noise or perhaps due to this absence of anisotropy. Finally, we find the orbitally averaged Fermi velocity for each pocket using the relation $v_F=\sqrt{2e\hbar F}/m^*$\, yielding $v_\gamma=$0.8$\times 10^{5}\, ms^{-1}$, $v_\beta=$1.3$\times 10^{5}\, ms^{-1}$ and $v_\alpha=$1.9$\times 10^{5}\, ms^{-1}$.

The $\gamma$ and $\beta$ frequencies observed here do not correspond to any of the Fermi surface pockets calculated for the non-magnetic state of BaFe$_2$As$_2$\cite{singh}. The $\alpha$ pocket is comparable in size to a calculated pocket centred at $\Gamma$, but the angle dependence reveals that the pocket is much more isotropic than that predicted in a non-magnetic calculation.  In agreement with the results in \sr122\, we conclude that a dramatic Fermi surface reconstruction has occurred in this compound. The Fermi surface pockets we observe are consistently larger than those in \sr122, but the effective masses are smaller, keeping the orbitally average velocities about the same \cite{sebastian}. In order to gain further insight into our data we perform band structure calculations including magnetic ordering which reconstruct the Fermi surface.\cite{bsfootnote} Our LDA/GGA calculations produce magnetic moments (1.67 and 1.97 $\mu_B$,
respectively) that are higher than the experimentally measured moment of
about 0.9 $\mu_B$ \cite{huang}  Because the Fermi surfaces have a
dependence on the magnitude of the moment, we suppressed it to the
experimental value using the well-known LDA+U methodology \cite{ldau} but
with a negative value for U.  Whereas LDA+U with U$>$0 generally increases
the magnetic tendencies of a system, the negative U has the opposite
effect and with a value of U=-0.54, we achieve a calculated magnetic
moment of 1.0 $\mu_B$.  The band structure with and without this technique
is shown in Fig. 3.  To check the validity of our methodology, we also increased the magnitude of the negative U until the magnetic moment was suppressed entirely and then compared the band structure and Fermi surfaces to their non-spin-polarized counterparts.  The agreement between Fermi surfaces was very good, though shifts of more than 100 meV could be found elsewhere in the energy spectrum.  Nonetheless, this partially suppressed moment calculation provides the best possible comparison to experiment. The results are summarized in Table I.

\begin{table}[tbp]
\begin{tabular}{|lcccccc|}
\hline
\hline
&$\mu$=1.6&&$\mu=1$&(Unshifted)&&(Shifted)\\
Orbit&$\%$BZ&$m_b$/$m_e$&$\%$BZ&$m_b$/$m_e$&$\%$BZ&$m_b$/$m_e$\\ \hline
1&0.01&-&0.06&0.1&0.3&0.47\\
2&0.15&0.2&0.29&0.27&0.7&0.55\\
3&3.1&1.6&3.34&1.16&1.63&0.7\\
4&6.7&2.5&4.6&1.95&-&-\\ \hline
$\gamma_s$ &&2.24&&2.83&&(mJ/molK$^2$)\\
\hline
\end{tabular}
\caption{Table of pocket size and effective mass expected to appear in magnetic QOs data, using a spin-polarized DFT calculation (described in text) for magnetic moment $\mu=1.6\mu_B$ and $\mu=1.0\mu_B$. The Sommerfeld coefficient $\gamma_s$ expected for each calculation is also given. Bands from the $\mu=1.0$ calculation have been shifted by different amounts to match the observed pockets, and the effective mass is recalculated.}
\label{bstable}
\end{table}

\begin{figure}
\includegraphics[width = 7.2cm ]{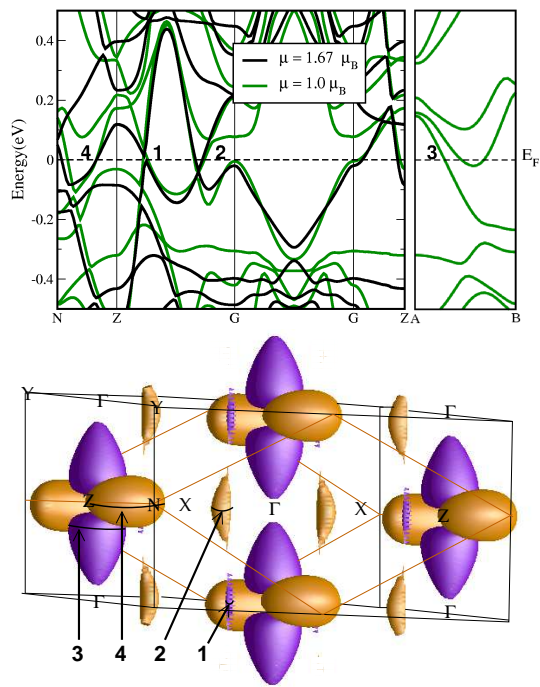}
\caption{(Top) Bandstructure for the reconstructed state of \ba122\, calculated for a magnetic moment of $\mu=1.6$ (black) and $\mu=1.0$ (green). Most of the pockets shown in Figure \ref{FS} are made up of multiple bands. The Fermi surfaces of BaFe$_2$As$_2$ in the  magnetic phase.  (Bottom) High symmetry directions corresponding to the band structure of Fig. \ref{FS} are shown in black.  The primitive Brillouin zone is shown in light (orange) non-orthogonal lines.  Extremal orbits $1$, $2$, $3$,and $4$ for (001) magnetic field are indicated.}  
\label{FS} 
\end{figure}

Four separate extremal orbits occur based on the calculated Fermi surfaces.  These are labelled in Fig. \ref{FS} as $1$ (hole), $2$ (electron), $3$ (hole), and $4$ (electron).  For the suppressed moment calculation, pocket $2$ has an area consistent with the measured F$_{\gamma}$.  The other calculated frequencies are either higher, $3$ and $4$, or much lower, $1$, than any observed frequencies.  However, it is common that calculations require an energy shift to achieve good agreement with experiment. The slightly oblate topology of $4$ is inconsistent with the observed angle dependence of any of the pockets, so we believe it unlikely that this is associated with the present QOs. To find an orbit area similar to the $\alpha$ pocket we find that a upward shift in E$_F$ of 60 meV would shrink pocket $3$ to a similar size with an effective mass of 0.7$m_e$, exhibiting a similar angle dependence to that observed. Both $1$ and $2$ have a topology consistent with the angle dependence of $\gamma$ and $\beta$ and the sharp pinching of the pockets at their extremities may explain the rapid loss of signal at large angles, due to the phase smearing associated with a higher Fermi surface curvature.  Similarly if $2$ is shifted upward by 33meV it comes into agreement with the $\beta$ orbit, with an effective mass of 0.55$m_e$. Finally $1$ is shifted by 38meV downward it comes into agreement with the $\gamma$ orbit, with an effective mass of 0.47$m_e$. We thus settle on identifying $\gamma$ with $1$, $\beta$ with $2$ and $\alpha$ with $3$ which is the same identification scheme than the one used for SrFe$_2$As$_2$\cite{sebastian}. With our assignment, the effective mass of $\alpha$ is renormalised by a factor of 1.7, $\beta$ by 1.1 and $\gamma$ is renormalised by 2 compared to DFT. All of the shifts are well within the $\sim$ 100 meV error incurred in suppressing the magnetic moment to zero, though the margin of error is expected to be somewhat less for the smaller shift to $\mu$=1.0$\mu_b$. 


The Sommerfeld coefficient has been extracted from heat capacity measured on these samples\cite{jiunhaw} and found to be $\gamma_S=6.1 mJ/molK^2$, consistent with Reference \onlinecite{dong}. The Sommerfeld coefficient calculated for the reconstructed Fermi surface is 2.83 mJ/mol K$^2$, as shown in Table I. This corresponds to a moderate renormalization of $\lambda$ = 1.1. This is consistent with the effective mass renormalization necessary to reconcile our QOs with DFT, which implies that the present calculations account for the observed Fermi surface. 
 

Taking the pockets we observe alone, we estimate a contribution of 1.73 mJ/molK$^2$. It is possible that the large Fermi cylinders observed by a number of ARPES measurements, centered at the $\Gamma$ point of the Brillouin zone\cite{yang,zhang,hsieh,zabolotnyy,ding} may account for the remainder. However, given that each 2D cylinder contributes $\sim$1.5 mJ/molK$^2$ per $m_e$, regardless of radius, the {\it maximum} effective mass available to
each cylinder is $\sim$1.4$m_e$, in contradiction to the ARPES results of Reference \cite{ding}. Furthermore, in order to reconcile the the existence of the large pockets with the antiferromagentic ground state of \ba122, an exotic nesting mechanism needs to be invoked\cite{zabolotnyy}. In contrast, the present measurements are consistent with a conventional SDW nesting picture.

In summary, we have measured QOs in \ba122\, and found small pockets which are in broad agreement with band structure calculations for an antiferromagnetic state. Our observations are consistent with a conventional spin -density wave picture which folds the bands of the non-magnetic state.

The authors would like to thank Nigel Hussey, Antony Carrington and Igor Mazin for useful comments on this work before publication. This work is supported by the Department of Energy, Office of Basic Energy Sciences under contract DE-AC02-76SF00515 and partly funded by EPSRC grant EP/F038836/1. Work performed at the NHMFL was primarily funded by NSF and the state of Florida.


\begin{thebibliography}{1}
\bibitem{errorfootnote} The observation of multiple orbits and the experimental signal/noise necessitates performing the Fourier analysis for extracting $m^*$ over a relatively wide range of B=20-60T. This combined with the large Dingle temperatures we observe is known to lead to a systematic underestimate of $m^*$. We estimate this to be $<20\%$ for both $m_\alpha^*$ and $m_\gamma^*$.
\bibitem{bsfootnote}Calculations were performed using an augmented plane wave plus local orbital (APW+lo) code, Wien2k \cite{wien}.  For the results presented in this work, we used the experimental lattice coordinates and ionic positions and the local density approximation (LDA) \cite{lda} to the exchange correlation potential.  We included the spin-orbit interaction in both LDA and the generalized gradient approximation (GGA) and the latter produces a higher magnetic moment, the electronic structure is sensitive to the magnitude of the moment and this, in turn, is very sensitive to the structure, specifically the height of the As atom above the Fe atom.
\end{thebibliography}
\end{document}